\documentclass{article}
\usepackage{graphicx} 
\usepackage{svg}
\usepackage[labelfont={bf, normalsize}]{subfig}
\usepackage{braket}
\usepackage[framemethod=tikz]{mdframed}
\usepackage[sorting=none,backend=biber]{biblatex}
\usepackage{amsmath}
\usepackage{amssymb}
\usepackage{xcolor}
\usepackage[top=1.5cm,bottom=2cm,left=1.75cm,right=1.75cm]{geometry}
\usepackage{hyperref}
\usepackage{mathtools}
\usepackage{mathrsfs}

\let\rket\ket
\let\rbra\bra

\newcommand{\ketbra}[2]{\rket{#1}\!\rbra{#2}}

\newcommand{\from}{\leftarrow}

\title{The CHSH Game, Tsirelson's Bound, and Causal Locality}
\author{Jacob A. Barandes$^{1}$\thanks{jacob\_barandes@harvard.edu}\ , Mahmudul Hasan$^{2}$\thanks{mahmudul.hasan@umassd.edu}\ ,\ and David Kagan$^{2}$\thanks{dkagan@umassd.edu}}

\date{\today}

\begin{document}

\maketitle

\begin{center}
\emph{\small{}$^{1}$Departments of Philosophy and Physics, Harvard University, 17 Oxford Street, Cambridge, MA 02138}\\
\emph{\small{}$^{2}$Department of Physics, University of Massachusetts
Dartmouth, North Dartmouth, MA 02747}
\par
\end{center}
\begin{abstract}
We reformulate the CHSH game in terms of indivisible stochastic processes. Using Barandes's stochastic-quantum correspondence and its associated definition of causal locality, we present a novel proof of the Tsirelson bound. In particular, we show that unlike the no-signaling principle alone, the postulates defining causally local, indivisible stochastic processes are precisely strong enough to allow for violations of the Bell inequality up to, but not beyond, the Tsirelson bound.
\end{abstract}

\section{Introduction}\label{sec:Intro}

Consider the CHSH inequality interpreted as arising from a coordination game \cite{Clauser1969,TsirelsonQuantumInfoProcessing}: Before any given round, two players, Alice and Bob, agree on a strategy. They are unable to communicate directly once a round has begun. During each round of the game, Alice randomizes a bit and then sends the resulting value, along with a second bit value of her choosing, to the referee, Charlie. Bob does the same. Alice knows the values of the two bits she sends to Charlie, and Bob likewise knows the values of his two bits, but neither can be aware of the other's randomized bit. To win the round, Alice and Bob's generated bits must somehow add up (modulo 2) to match the product of their random bits.

Under the assumptions of various versions of Bell's theorem \cite{Bell1964,Bell1975,Bell2004}, Alice and Bob should not be able to win more than 3/4ths of the rounds, even if they are allowed to share correlated bits as resources. However, if the players share a suitably prepared entangled pair of qubits, then they can exceed the limitations of Bell's theorem and saturate Tsirelson's bound, winning $\cos^2(\pi/8) \approx 85.4\%$ of the time \cite{Cirelson1980}.

Known physical systems, whether classical or quantum, obey the no-signaling principle, which is roughly that probabilities for local events depend only on locally available data, thereby ruling out the possibility of superluminal signals \cite{Ghirardi1980,Jordan1983}. In \cite{Popescu1994}, Popescu and Rohrlich studied whether imposing no-signaling was sufficient to explain the existence of the Tsirelson bound. Utilizing a construction by Rastall \cite{Rastall1985}, Popescu and Rohrlich demonstrated the existence of no-signaling models that maximally exceed Tsirelson's bound. If such resources exist in nature, then they would allow Alice and Bob to win the CHSH game 100\% of the time.

Why then does our world appear to realize the intermediate range, violating Bell's inequality while respecting Tsirelson's bound? In \cite{Pawlowski2010}, the authors put forward their principle of information causality, which states that local systems can gain only as much information about a remote system as is communicated to them. According to \cite{Pawlowski2010,Pawlowski2021}, information causality can indeed be used to rule out no-signaling theories that exceed Tsirelson's bound, though to arrive at the precise bound requires considering protocols that involve either a large number of bits or noisy channels.

Is there another way to understand why our world allows correlations stronger than Bell's, but not the theoretical maximum consistent with no-signaling? The stochastic-quantum correspondence \cite{SQCorrespondence,SQTheorem,Barandes2025} and the principle of causal locality formulated in \cite{SQCausalLocality} suggest an alternative. Consider the following postulates:

\begin{enumerate}
    \item The existence of system configurations: A physical system's configurations form a set,\footnote{We do not assume a richer mathematical structure, such as a Hilbert space. The case of systems with finite sets of configurations is sufficient for our purposes.} and the system is in one of its possible configurations at any given instant. The configuration set of a composite system is the Cartesian product of the corresponding subsystem configuration sets.
    \item A stochastic description of time-directed dynamics: The laws governing a physical system's evolution are described via time-directed conditional probabilities relating probabilities of configurations at a given conditioning time to probabilities of configurations at target times. The dynamics are generically non-Markovian. We will often take the conditioning time to be an initial time preceding a later target time, but there need not be any fundamental breaking of time-reversal symmetry or fundamental arrow of time. Subsystems are governed by marginal distributions derived from the laws of their parent systems.
    \item Causal locality: Physical influences as described by a system's stochastic dynamics cannot propagate faster than the speed of light. That is, the stochastic dynamics of a localized system during a specific time interval cannot be conditionally dependent on any other systems beyond the range of light-like signals during that time interval.
\end{enumerate}

In \cite{SQCorrespondence,SQTheorem,Barandes2025}, Barandes showed that the first two postulates are sufficient to describe all known physical systems, including quantum systems. Here, we illustrate how stochastic systems that respect the third postulate can violate Bell inequalities up to Tsirelson's bound. Note the implication that any non-signaling scenario involving Tsirelson-violating correlations, such as PR boxes, cannot be modeled using the indivisible stochastic systems of \cite{SQCorrespondence}.

Section \ref{sec:CHSHReview} reviews the CHSH game and the no-signaling (NS) box framework that demonstrates the existence of probabilistic strategies that win the game all the way up to the theoretical maximum, assuming suitable non-signaling resources shared by Alice and Bob. Section \ref{sec:SQandCausalLocality} outlines the stochastic-quantum correspondence and some of the key ideas for the definition of causal locality that we use in this work.  In Section \ref{sec:StochasticCHSH}, we recast an arbitrary round of the CHSH game in stochastic terms. We then prove that adopting causal locality yields the Tsirelson bound. In Section \ref{sec:Conclusion}, we consider some questions about the nature of causality and locality in quantum theory versus the stochastic picture, as well as suggesting some directions for future research.

\section{Review of the CHSH Game}\label{sec:CHSHReview}

\subsection{Setup}\label{sec:CHSH-setup}

In a given round of the CHSH game, two players, Alice and Bob, are spacelike separated. Alice randomizes a bit $x\in \{0,1\}$, which she sends to Charlie, along with another bit $q\in\{0,1\}$ of her choosing. Alice is aware of both bit values. Bob follows the same procedure, randomizing $y\in\{0,1\}$ and choosing $r\in\{0,1\}$, sending both to Charlie. Charlie tests whether the condition $q\oplus r = x y$ holds, where the operation $\oplus$ denotes addition modulo 2. If Alice and Bob win, they are awarded a point; if they lose, they are docked a point.

Alice and Bob share a bipartite resource, prepared by Charlie in advance according to instructions that Alice and Bob agreed to before the start of the game. Charlie sends Alice one part of the resource and Bob the other. Only one bit of information is extractable from each part of the resource.

Let $P(q,r,x,y)$ denote the probability that Alice and Bob return the bit values $q,r,x$, and $y$. Their expected score in any given round of the game is
\begin{equation}\label{eq:CHSH-score-0}
    \begin{split} 
        \langle \textrm{score} \rangle  &=  \sum_{\mathclap{x,y,q,r} } \big(P(q,r,x,y)\delta(q\oplus r = xy) - P(q,r,x,y)\delta(q\oplus r \neq xy)\big) \\
        & = \sum_{\mathclap{x,y,q,r} } (-1)^{xy-q\oplus r}P(q,r|x,y)P(x,y) \\
        & = \sum_{\mathclap{x,y} } (-1)^{xy}\big(P(0,0|x,y)-P(0,1|x,y)-P(1,0|x,y)+P(1,1|x,y)\big)P(x,y),
    \end{split}
\end{equation}
where $P(x,y)$ is the marginal joint probability of the bit value pair $(x,y)$ and $P(q,r|x,y) \equiv P(q,r,x,y)/P(x,y)$. The function $\delta(\textrm{condition})=1$ if the condition in its argument is true, and is 0 otherwise. We can re-express the score by substituting $P(1,1|x,y)=1-P(0,0|x,y)-P(0,1|x,y)-P(1,0|x,y)$ into the last expression, yielding
\begin{equation}\label{eq:CHSH-score-1}
    \begin{split} 
        \langle \textrm{score} \rangle
        & = \sum_{\mathclap{x,y} } (-1)^{xy} \big(1-2P(0,1|x,y)-2P(1,0|x,y)\big)P(x,y) \\
        & = \sum_{\mathclap{x,y}} (-1)^{xy} P(x,y)-2 \sum_{\mathclap{x,y} } (-1)^{xy} \big(P(0,1|x,y)+P(1,0|x,y)\big)P(x,y) \\
        & = 1-2P(x=1,y=1)-2\sum_{\mathclap{x,y} } (-1)^{xy}\big(P(0,1|x,y)+P(1,0|x,y)\big)P(x,y),
    \end{split}
\end{equation}
where the last equality in (\ref{eq:CHSH-score-1}) follows from
\begin{equation}
    \sum_{x,y} (-1)^{xy} P(x,y) = P(0,0) + P(0,1) + P(1,0) - P(1,1) = 1 - 2P(1,1).
\end{equation}
If Alice and Bob generate the values $x$ and $y$ according to a uniform distribution, then $P(x,y)=1/4$. Thus,
\begin{equation}\label{eq:CHSH-score-2}
    \braket{\textrm{score}} = \frac{1}{2}\Bigg(1-\sum_{x,y}(-1)^{xy}\big(P(0,1|x,y) + P(1,0|x,y)\big)\Bigg).
\end{equation}
Equivalently, from (\ref{eq:CHSH-score-0}), we have
\begin{equation}\label{eq:CHSH-score-3}
    \braket{\textrm{score}} = \frac{1}{4}\sum_{\mathclap{x,y} } (-1)^{xy}\big(P(0,0|x,y)-P(0,1|x,y)-P(1,0|x,y)+P(1,1|x,y)\big).
\end{equation}

\subsection{No-Signaling and NS Boxes}

The conditional probability distribution $P(q,r|x,y)$ describes all possible correlations between Alice and Bob's response bits, given the pair of input bits $(x,y)$ they randomly generate and their use of the shared resource. Note that at this stage, $P(q,r|x,y)$ is merely descriptive, and does not commit us to any particular underlying dynamical picture. The no-signaling principle is a physical criterion that is meant to constrain $P(q,r|x,y)$, and thus indirectly encode some physical information in the correlations the conditional probabilities describe. These no-signaling correlations satisfy
\begin{equation}\label{eq:CHSH-noSignaling}
    P(q|x,y) = P(q|x),\qquad P(r|x,y)=P(r|y),
\end{equation}
where the marginal probabilities $P(q|x,y)$ and $P(r|x,y)$ are defined by
\begin{equation}
    P(q|x,y) \equiv \sum_r P(q,r|x,y),\qquad P(r|x,y) \equiv \sum_q P(q,r|x,y).
\end{equation}
The no-signaling constraints describe the intuition that outcomes seen only locally by either Alice or Bob should only be affected by local influences. In the CHSH context, Alice's outcomes for $q$ should only be correlated with the bit value $x$, and, similarly, Bob's outcomes for $r$ should only be correlated with $y$.

Following \cite{Pawlowski2010,MasanesSignaling2006}, we can characterize non-signaling resources using the family of distributions referred to as no-signaling (NS) boxes,
\begin{equation}\label{eq:NSBox}
    P_{NS}(q,r|x,y) = \frac{1}{4}\big(1+(-1)^{xy-q\oplus r}E\big).
\end{equation}
Notice that this distribution is locally unbiased in that Alice and Bob each have marginal probabilities of $1/2$ for any of the outcomes $q=0,1$ and $r=0,1$, respectively. In particular, it satisfies the no-signaling conditions (\ref{eq:CHSH-noSignaling}). Substituting (\ref{eq:NSBox}) into (\ref{eq:CHSH-score-3}) shows that the parameter $E$ is equivalent to $\braket{\textrm{score}}$.  The case $E=0$ corresponds to Alice and Bob ignoring their random bit values, $x$ and $y$, respectively, and simply selecting $q$ and $r$ randomly with the equivalent of fair coin tosses. When $E=1/2$, Alice and Bob have chosen a strategy that allows them to win 3/4ths of the time, saturating Bell's inequality. The quantum upper limit corresponds to $E=1/\sqrt{2}$, while the `PR box' construction of \cite{Popescu1994,Rastall1985} corresponds to $E=1$. 

The no-signaling principle places no restrictions on the score parameter $E$ in (\ref{eq:NSBox}), and thus lacks the explanatory resources to single out the Tsirelson bound.

\section{The Stochastic-Quantum Correspondence and Causal Locality}\label{sec:SQandCausalLocality}

\subsection{Indivisible Stochastic Systems}

Let $\mathcal{C}$ be the set of configurations of a system $Q$. For simplicity, we assume that $\mathcal{C}$ is finite, and let  $q_t=1,2,\ldots ,N$ label the possible configuration of $Q$ at time $t$. When we refer to a specific time $t_n$ that carries an index of its own, we employ the notation $q_n \equiv q_{t_n}$. We assume at least one conditioning time $t_0$ at which $Q$ has no correlations with any other systems. 

Conditional probabilities $p(q_t|q_0)$ encode stochastic dynamics. Given an initial standalone probability distribution over configurations $p(q_0) \equiv p(q_0;t_0)$, we may define a joint probability distribution,
\begin{equation}
    p(q_t,q_0) = p(q_t|q_0) p(q_0),
\end{equation}
and marginalize over the configuration at the initial time to arrive at the standalone probability for configuration $q$ at time $t$,
\begin{equation}
   \label{eq:StochasticEvo} 
   p(q_t) = \sum_{q_0} p(q_t|q_0) p(q_0).
\end{equation}
The formalism works for either $t < t_0$ or $t>t_0$, but, for specificity, we will assume that $t_0$ is the earliest time under consideration at which our system is uncorrelated with anything else and that all other times in the process occur later.

Generically, systems evolve indivisibly, except for specific instants, including $t_0$, called division events or division times. Indivisible evolution means that the system's configurations cannot be tracked, even in principle, which is reflected by the lack of any assignment of probabilities to system trajectories. By contrast, a divisible stochastic process during the time interval $t_0\to t$  has division times at each instant within the given time interval, up to the given level of precision.

We define a stochastic matrix $\Gamma(t\from t_0)$ with components $\Gamma_{q_t q_0}(t\from t_0) \equiv p(q_t|q_0)$. An indivisible stochastic process is called divisible at a time $t_1$ if the stochastic matrix factorizes at that time,
\begin{equation}\label{eq:Background:MarkovianDivision-GammaForm}
    \Gamma(t\from t_0) = \Gamma(t\from t_1) \Gamma(t_1 \from t_0),
\end{equation}
where $\Gamma(t\from t_1)$ is a well-defined stochastic matrix of its own. In that case, we axiomatically take the matrix elements of $\Gamma(t\from t_1)$ to be conditional probabilities so that, given the standalone distribution at time $t_1>t_0$,
\begin{equation}
    p(q_1) = \sum_{q_0} p(q_1|q_0) p(q_0),
\end{equation}
we may consistently write
\begin{equation}\label{eq:Background:MarkovianDivision-ConditionalForm}
     p(q_t) = \sum_{q_1} p(q_t|q_1) p(q_1),
\end{equation}
where $t>t_1$. As with $t_0$, the time $t_1$ is called a division event or division time. Correlations between the system $Q$ and other systems are effectively washed out during a division event. Division events are often only approximate or effective, such as when they arise due to interactions between a system and a larger environment. We adopt a pragmatic perspective that all dynamical probabilistic statements are defined up to some given level of precision, which demarcates the validity of approximating a process as a division event.

\subsubsection{Composite Systems}

As stipulated in the first postulate in Section \ref{sec:Intro}, a configuration of a composite system $QR$ can be expressed as an ordered pair $(q_t,r_t)$ of configurations $q_t$ and $r_t$ of the individual systems $Q$ and $R$, respectively. Thus, the configurations of $QR$ form a Cartesian product, $\mathcal{C}_{QR}=\mathcal{C}_Q \times \mathcal{C}_R$,  where $\mathcal{C}_Q$ and $\mathcal{C}_R$ are the configuration sets of the systems $Q$ and $R$, respectively. Given an initial division event at $t_0$, the joint stochastic dynamics for $QR$ are defined by conditional probabilities of the form $p(q_t,r_t|q_0,r_0)$ that describe the evolution of the initial probability distribution to later times $t$,
\begin{equation}
    p(q_t,r_t) = \sum_{q_0,r_0} p(q_t,r_t|q_0,r_0) p(q_0,r_0).
\end{equation}
Marginalizing yields mixed stochastic dynamics for $Q$ and $R$,
\begin{equation}
    p(q_t|q_0,r_0) = \sum_{r_t} p(q_t,r_t|q_0,r_0),\qquad p(r_t|q_0,r_0) = \sum_{q_t} p(q_t,r_t|q_0,r_0).
\end{equation}
If the conditional probabilities factorize,
\begin{equation}\label{eq:Background:DivisionConditionalFactorization}
    p(q_t,r_t|q_0,r_0) = p(q_t|q_0)p(r_t|r_0),
\end{equation}
then we say that the subsystems $Q$ and $R$ do not interact during the interval $t_0\to t$. If $Q$ and $R$ interact at a later time, then their composite stochastic dynamics will cease to factorize until a subsequent division event. Note that interactions are an example of the broader notion of causal influence defined in \cite{SQCausalLocality}, and described in more detail in Section \ref{sec:CausalInfluence}.

\subsection{The Stochastic-Quantum Correspondence}

\subsubsection*{Trace Formula for Conditional Probabilities}

A system's stochastic dynamics can be (non-uniquely) represented in terms of complex-valued elements of a matrix $\Theta(t\from t_0)$,
\begin{equation}
    \Theta_{q_tq_0}(t\from t_0) = e^{i\theta_{q_tq_0}(t)}\sqrt{p(q_t|q_0)},\qquad p(q_t|q_0) = |\Theta_{q_tq_0}(t\from t_0)|^2,
\end{equation}
where $\theta_{q_tq_0}(t)$ is an arbitrary phase. For the moment, let us consider the case where $\Theta(t\from t_0) = U(t\from t_0)$ is a unitary matrix. In that case, the corresponding matrix $\Gamma(t\from t_0)$ of conditional probabilities is called unistochastic\cite{Horn1954}. The matrix  $U(t\from t_0)$ acts linearly on a Hilbert space $H_\mathcal{C}$ of dimension $N$. The configurations of the system define a preferred basis, $\{\ket{q}\}_q$, for $H_\mathcal{C}$. However, $U(t\from t_0)$ generically carries configuration-basis vectors to other vectors that are superpositions of configuration-basis vectors. These are not regarded as physical configurations of the system, but rather as a way to characterize the behavior of the system when it undergoes indivisible evolution.

Returning to the case of generic $\Theta(t\from t_0)$, we let $\mathcal{P}_q=\ketbra{q}{q}$ be a projection matrix onto the configuration-basis vector $\ket{q}$. The stochastic dynamic probabilities can now be expressed via the trace
\begin{equation}\label{eq:Dictionary}
    p(q_t|q_0) = \textrm{Tr}\big[\mathcal{P}_{q_t} \Theta(t\from t_0) \mathcal{P}_{q_0} \Theta^\dagger(t\from t_0)\big].
\end{equation}
The relation (\ref{eq:Dictionary}) is the dictionary of the stochastic-quantum correspondence.

\subsubsection*{Unistochastic Dynamics}

Any indivisible stochastic system's quantum description can be dilated so that the enlarged system evolves according to a unitary time-evolution matrix $U(t\from t_0)$ with matrix elements $U_{q_t q_0}$, as demonstrated in \cite{SQTheorem}. Processes that can be represented unitarily are called unistochastic processes. Their conditional probabilities satisfy
\begin{equation}
    p(q_t|q_0) = \textrm{Tr}\big[\mathcal{P}_{q_t} U(t\from t_0) \mathcal{P}_{q_0} U^\dagger(t\from t_0)\big] = \big|\!\braket{q_t|U(t\from t_0)|q_0}\big|^2.
\end{equation}

Unitarity permits us to define $U(t\from t')\equiv U(t\from t_0) U^\dagger(t'\from t_0)$ for any $t'<t$, which implies that $U(t\from t_0)=U(t\leftarrow t')U(t'\from t_0)$, and hence
\begin{eqnarray}
    \label{eq:Background:unistochasticFactors}
    p(q_t|q_0) &=& \sum_{q',q''} U_{q_t q'}^* U_{q'q_0}^*  U_{q_t q''} U_{q''q_0} \nonumber \\
    &=& \sum_{q'} \big|U_{q_t q'}\big|^2\big| U_{q'q_0}\big|^2+\sum_{q'\neq q''} U_{q_t q'}^* U_{q'q_0}^*  U_{q_t q''} U_{q''q_0}\nonumber\\
    &=& \Gamma'(t\from t')\Gamma(t'\from t_0) + \textrm{qCor}(t\from t'\from t_0).
\end{eqnarray}
Here the index $q'$ corresponds to the intermediate configuration label at time $t'$, and the matrix $\textrm{qCor}(t\from t'\from t_0)$ captures the discrepancy between actual indivisible unistochastic evolution and fictitious unistochastic evolution with a division event at $t'$. Such discrepancies are typically referred to as quantum interference effects. The stochastic-quantum correspondence thus situates quantum interference within the broader context of discrepancies between exact non-Markovian processes and any Markovian approximation.

\subsubsection*{Division Events and Decoherence}

Stochastic evolution is generically indivisible. Given a system $Q$, let $\mathcal{T}_\textrm{div}$ denote the set of division times where
\begin{equation}
    \Gamma(t\from t_0) = \Gamma(t\from t_n)\Gamma(t_n\from t_{n-1})\cdots \Gamma(t_1\from t_0),\qquad t_0,t_1,\ldots,t_n \in \mathcal{T}_\textrm{div},
\end{equation}
and $t_0<t_1<\ldots<t_n<t$. The stochastic-quantum correspondence associates effective division events with environmental decoherence \cite{SQCorrespondence}.\footnote{Exact division events can also occur. See the example given in \cite{SQCorrespondence}.} Such decoherence can be modeled in terms of dephasing channels $\mathcal{D}_i$ that remove density-matrix coherence terms at the times $t_i \in \mathcal{T}_\textrm{div}$ as expressed in the configuration basis. We refer to these as division channels.

\subsection{Causal Influence and Causal Locality}\label{sec:CausalInfluence}

Consider two systems $Q$ and $R$. Following \cite{SQCausalLocality}, we say that $R$ does \textit{not} causally influence $Q$ during the time interval starting with the initial division time $t_0$ until the later target time $t$ if
\begin{equation}\label{eq:CausalInfluence}
p(q_t|q_0,r_0) = p(q_t|q_0),
\end{equation}
where $p(q_t|q_0,r_0)$ is the marginal conditional probability distribution derived from the joint conditional probability distribution $p(q_t,r_t|q_0,r_0)$.  Otherwise, $R$ does causally influence $Q$. Note that no external observers or measurements are invoked in this approach to causality. The laws formulated as conditional probabilities themselves provide the causal description in a non-interventionist variant of causal Bayesian network theory \cite{Pearl2009}.

If $Q$ and $R$ are causally independent, then their marginal conditional probability distributions satisfy
\begin{eqnarray}\label{eq:CausalIndependence}
    p(q_t\vert q_0,r_0) &=&  p(q_t\vert q_0), \\
    p(r_t\vert q_0,r_0) &=&  p(r_t\vert r_0).
\end{eqnarray}
The conditions (\ref{eq:CausalIndependence}) are weaker than the factorization condition (\ref{eq:Background:DivisionConditionalFactorization}) that characterizes non-interaction. Thus, interaction is a special case of causal influence.

Suppose that $Q$ and $R$ are spatially separated. They form a causally local system if they are causally independent when $c \cdot(t-t_0)$ is shorter than their separation distance, with $c$ being the speed of light.   Notice how the stochastic picture, in postulating a level of physical reality beneath the Hilbert spaces of standard quantum theory, makes it possible to formulate this new criterion of causal locality, which will turn out to be stronger than the no-signaling principle and will make it possible to derive the Tsirelson bound.

\section{Stochastic Description of the CHSH Game}\label{sec:StochasticCHSH}

Consider the overarching system made up of Alice $A$, Bob $B$, Charlie $C$, particle $Q$, particle $R$, and their local environment $E$. Alice, Bob, and Charlie are monitored by the environment nearly continuously, and so their local stochastic dynamics are effectively divisible. Particles $Q$ and $R$ may be shielded from the environment, and may thus undergo indivisible unistochastic evolution.

Alice, Bob, and Charlie's configurations at any given time $t$ are denoted by $a_t(\ldots),b_t(\ldots)$, and $c_t(\ldots)$, respectively, where `$\ldots$' represent details of a particular configuration, potentially including memories of past configurations. Alice and Bob are each able to send Charlie two bits of information per round. Charlie can also transmit $Q$ and $R$ to Alice and Bob, respectively. Finally, Charlie, Alice, and Bob can perform local operations that are described by indivisible stochastic processes on systems $Q$ and $R$.

 \begin{figure}
     \centering
     \includegraphics[width=0.4\linewidth]{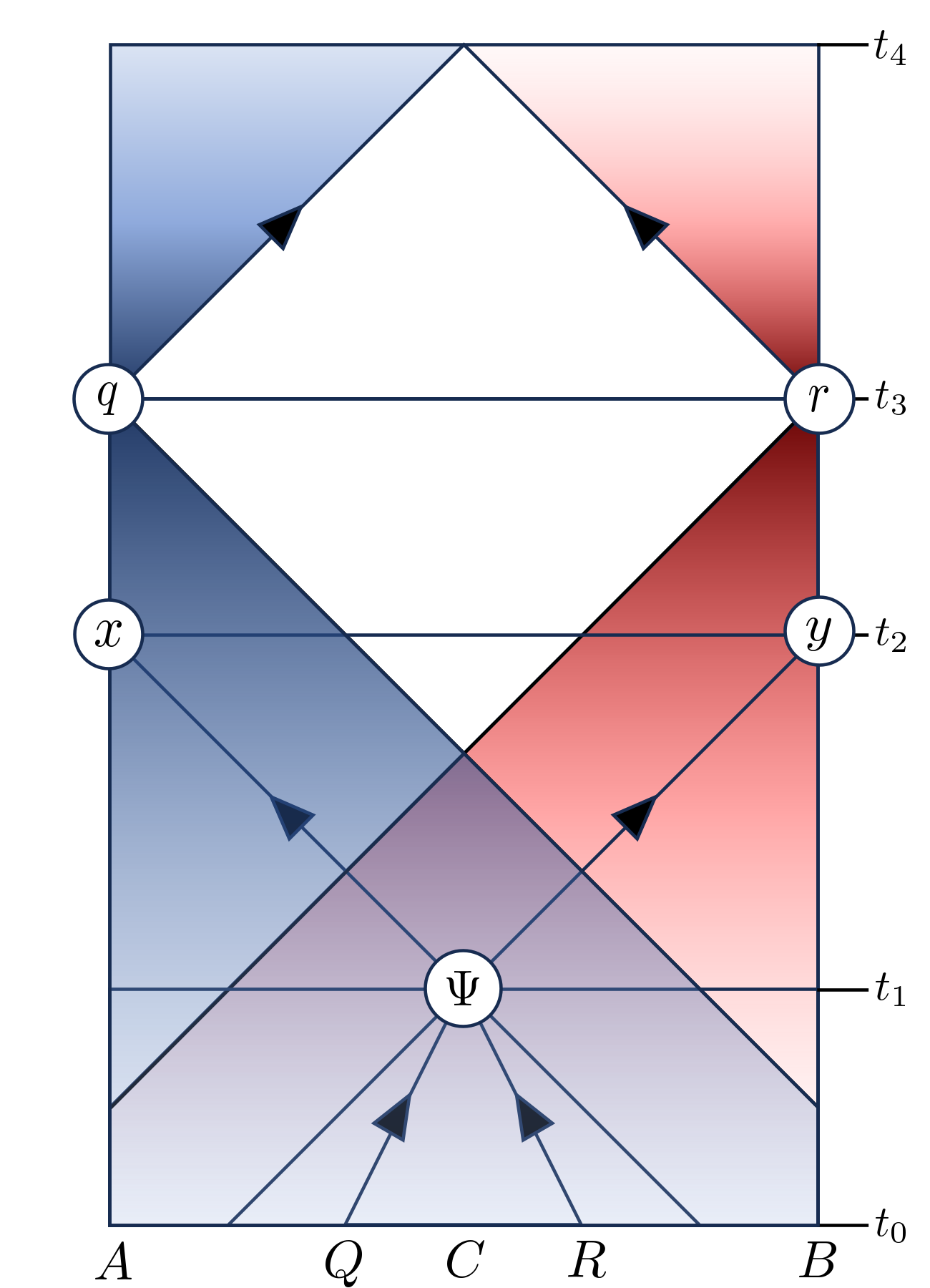}
     \caption{A spacetime diagram depicting one round of the CHSH game.}
     \label{fig:CHSH-diagram}
 \end{figure}

Let $t_0$ denote a division event for all of the systems initiating one round of the game. At this time, Charlie has particles $Q$ and $R$ in configurations $q_0$ and $r_0$, respectively. During the interval $t_0\to t_1$, Charlie implements the instructions given to him by Alice and Bob before the round began. At $t_1$, Charlie's configuration will include information about the preparation of the composite system $QR$, which we denote $c_1(\Psi,\ldots)$. The preparation may involve joining $Q$ and $R$ in an indivisible process, in which case the label $\Psi$ in Charlie's memory is correlated with a process rather than with any particular configuration $(q_1,r_1)$ of the joint system $QR$.

Charlie sends $Q$ to Alice and $R$ to Bob at time $t_1$, with the particles arriving at time $t_2$. On receiving their particles, Alice and Bob generate the random bits $x$ and $y$, respectively. The bit values are reflected in Alice and Bob's configurations as $a_2(x,\ldots)$ and $b_2(y,\ldots)$. Alice and Bob perform local operations on $Q$ and $R$, respectively, during the interval $t_2\to t_3$. Note that Alice and Bob's choices of local operations may be influenced by the values of $x$ and $y$, respectively, as part of their prearranged strategy. Causal locality plays an important role here, ensuring that Alice's choice of operation depends solely on $x$ and Bob's solely on $y$.

At time $t_3$, Alice and Bob measure the configurations of $Q$ and $R$ to be $q$ and $r$, respectively. Alice and Bob will choose their discretionary bits to be their measurement-outcome bits $q$ and $r$, respectively. They communicate all of their bit values $x,y,q,$ and $r$ to Charlie, who receives the information at time $t_4$ and whose configuration is now of the form $c_4(q,r,x,y,\Psi,\ldots)$. Charlie is thus able to check the win condition for the round and keep a tally of the score in memory. 

The overall process described above results in a hybrid stochastic distribution for $q$ and $r$ at times $t\geq t_3$ conditioned on Alice's knowledge $a_2(x)$ of $x$, Bob's knowledge $b_2(y)$ of $y$, and Charlie's preparation $c_1(\Psi)$,
\begin{equation}\label{eq:StochasticDistributionNotation}
    p_{xy}(q,r|\Psi) \equiv p\big(q,r;t|a_2(x),b_2(y),c_1(\Psi)\big),\quad t\geq t_3,
\end{equation}
where $a_2(x)$, $b_2(y)$, and $c_1(\Psi)$ lie along a single Cauchy hypersurface. The left-hand side of (\ref{eq:StochasticDistributionNotation}) introduces a notational shorthand, while the right-hand side makes explicit the conditional dependencies inherent in the stochastic process we are describing. We use lower-case `$p$' to distinguish the stochastic distribution (\ref{eq:StochasticDistributionNotation}) above from the generic distribution $P(q,r|x,y)$ introduced in Section \ref{sec:CHSH-setup}.

\subsection{The Role of Causal Locality}

In the CHSH context, causal locality implies further that
\begin{eqnarray}
    p_{xy}\big(q\vert\Psi\big) &\equiv& p_{x}\big(q\vert\Psi\big) = p\big(q;t|a_2(x),c_1(\Psi)\big) \\
      p_{xy}\big(r\vert\Psi\big) &\equiv& p_{y}\big(r\vert\Psi\big) = p\big(r;t|b_2(y),c_1(\Psi)\big).
\end{eqnarray}
Causal locality thus defines what it means for Alice and Bob's operations to be local: Configuration $r$ is outside of the future light cone of random bit $x$, and thus cannot be causally influenced by $x$. Similarly, configuration $q$ is outside of the future light cone of random bit $y$, and thus cannot be causally influenced by $y$. However, the marginal probabilities for outcomes $q$ and $r$ both generically depend on the result of the preparation $\Psi$ at time $t_1$, due to that indivisible process being in both of their causal pasts.

 \begin{figure}
     \centering
     \includegraphics[width=0.4\linewidth]{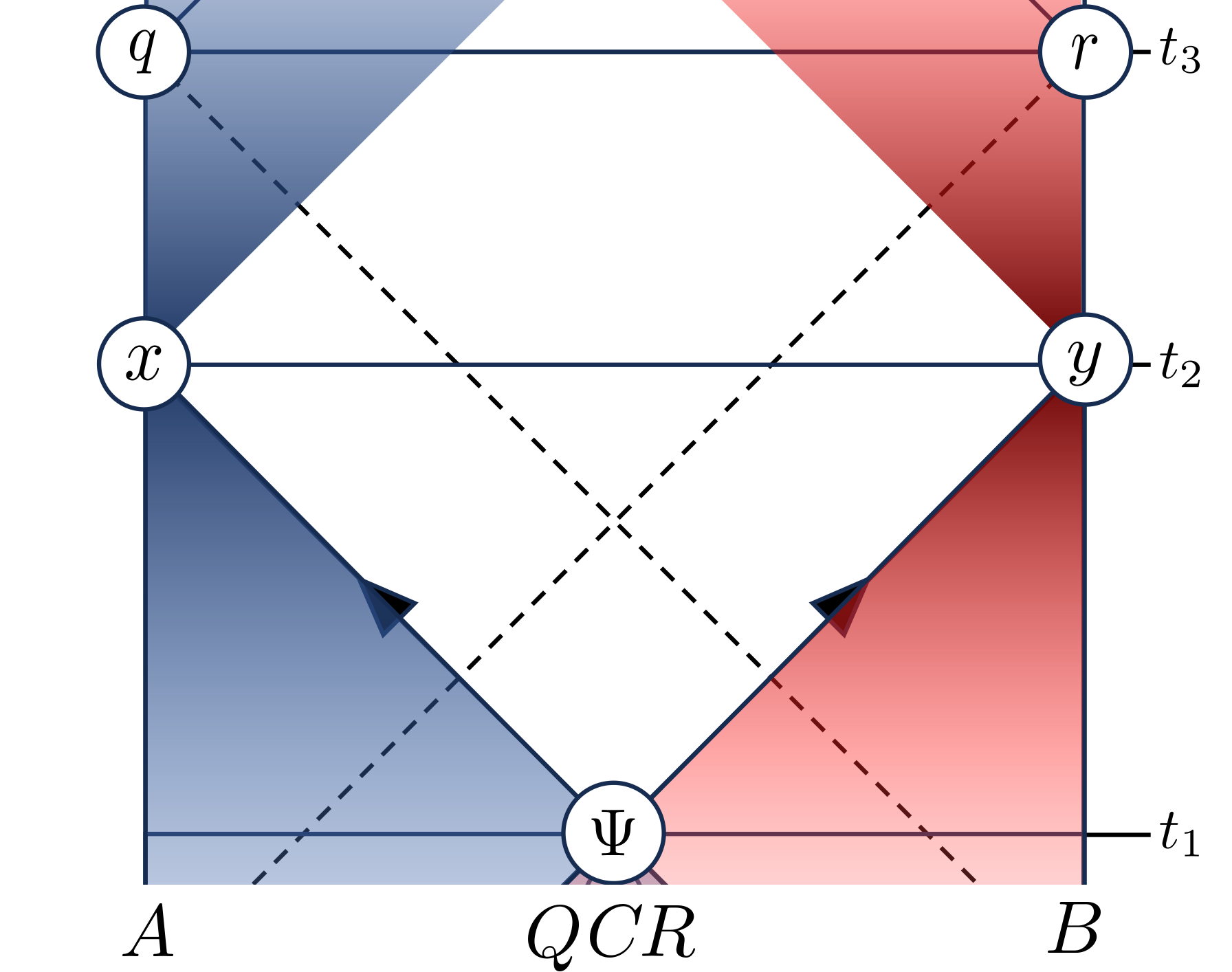}
         \caption{At time $t_3$, the configuration $q$ of $Q$ may depend on the random bit $x$ as it lies in the blue-shaded future light cone in the diagram. Similarly, the configuration $r$ of $R$ at time $t_3$ lies in the red-shaded future light cone of random bit $y$, and may therefore depend on $y$. Both configurations $q$ and $r$ will generically depend on the preparation $\Psi$ as it lies in the overlap of their past light cones under the intersection of the dotted lines in the diagram. However at $t_3$, causal locality implies that $r$ cannot depend on $x$ and $q$ cannot depend on $y$ as each lies outside of the future light cones of $x$ and $y$, respectively.}
     \label{fig:CausalLocalityImplications}
 \end{figure}

The no-signaling principle by itself does not constrain the nature of the processes underlying the descriptive joint conditional probability distribution $P(q,r|x,y)$ of Section \ref{sec:CHSHReview}. By contrast, causal locality is more stringent, requiring a stochastic picture with time-directed conditional probabilities that describe the underlying laws governing the system.

At this stage, we claim
\begin{quote}
\textbf{Tsirelson's bound:} The stochastic-quantum correspondence implies that the CHSH game score associated with a causally local unistochastic process given by $p_{xy}(q,r|\Psi)$ can be no greater than $1/\sqrt{2}$. Equivalently, the total win probability cannot exceed $(1+\sqrt{2})/2\sqrt{2}$.
\end{quote}
 \subsection{Proof}

Suppose that $QR$ evolves unistochastically up until Alice and Bob's measurements at time $t_3$. According to the stochastic-quantum dictionary established in \cite{SQCorrespondence}, we have
\begin{equation}\label{eq:quantumConditional}
    \begin{split}
    p_{xy}\big(q,r\vert\Psi\big) &= \textrm{Tr}\Big[\big(\mathcal{P}_q \otimes \mathcal{P}_r\big)\mathbf{(\mathcal{AB})}_{xy}\,\mathbf{\mathcal{C}}_\Psi \ket{q_0,r_0}\bra{q_0,r_0}\,\mathbf{\mathcal{C}}^\dagger_{\Psi}\mathbf{(\mathcal{AB})}_{xy}^\dagger\Big] \\
    &=\textrm{Tr}\Big[\big(\mathcal{P}_q \otimes \mathcal{P}_r\big)\mathbf{(\mathcal{AB})}_{xy} \ket{\Psi}\bra{\Psi}\mathbf{(\mathcal{AB})}_{xy}^\dagger\Big],
    \end{split}
\end{equation}
where $\mathcal{P}_q=\ketbra{q}{q}$, $\mathcal{P}_r=\ketbra{r}{r}$, $\mathbf{(\mathcal{AB})}_{xy}$ is a unitary representation of Alice and Bob's joint operations on $QR$, and $\mathbf{\mathcal{C}}_\Psi$ is a unitary representation of Charlie's preparation of $QR$. 

Using properties of the trace, we may re-express (\ref{eq:quantumConditional}) as,
\begin{equation}\label{eq:quantumConditional2}
    p_{xy}\big(q,r\vert\Psi\big) = \braket{\Psi\big\vert\mathbf{(\mathcal{AB})}_{xy}^\dagger \big(\mathcal{P}_q \otimes \mathcal{P}_r\big)\mathbf{(\mathcal{AB})}_{xy}\big\vert {\Psi}}.
\end{equation}
Causal locality requires Alice and Bob's joint operation to factorize: $\mathbf{(\mathcal{AB})}_{xy} = \mathcal{A}_x  \otimes \mathcal{B}_y$. Therefore,
\begin{equation}\label{eq:quantumConditionalFactored}
    p_{xy}\big(q,r\vert\Psi\big) = \braket{\Psi\big\vert\mathcal{A}_x^\dagger \mathcal{P}_q \mathcal{A}_x \otimes \mathcal{B}_y^\dagger \mathcal{P}_r \mathcal{B}_y \big\vert {\Psi}}.
\end{equation}
We define the Hermitian operators
\begin{equation}\label{eq:ABHermitian}
    A_x^q \equiv \mathcal{A}_x^\dagger \mathcal{P}_q \mathcal{A}_x\otimes \mathbb{I},\qquad B_y^r \equiv \mathbb{I}\otimes \mathcal{B}_y^\dagger \mathcal{P}_r \mathcal{B}_y.
\end{equation}
We calculate the CHSH score for this setup by substituting $P(q,r|x,y)=p_{xy}(q,r|\Psi)$ into the formula (\ref{eq:CHSH-score-3}),
\begin{eqnarray}\label{eq:QuantumScore}
    \braket{\textrm{score}} &=& \frac{1}{4}\sum_{\mathclap{x,y} } (-1)^{xy}\big(p_{xy}(0,0|\Psi)-p_{xy}(0,1|\Psi)-p_{xy}(1,0|\Psi)+p_{xy}(1,1|\Psi)\big) \\
    &=& \frac{1}{4}  \braket{\Psi\big \vert \sum_{x,y} (-1)^{xy}(A^0_x B^0_y - A^0_x B^1_y - A^1_x B^0_y + A^1_x B^1_y) \big\vert \Psi }  \nonumber \\
    &=&  \frac{1}{4}\braket{\Psi\big \vert \sum_{x,y} (-1)^{xy} \mathscr{A}_x \mathscr{B}_y\big\vert \Psi } \nonumber \\
    &=& \frac{1}{4}\braket{\Psi\big \vert \mathscr{A}_0 \big(\mathscr{B}_0 + \mathscr{B}_1\big) + \mathscr{A}_1 \big(\mathscr{B}_0 - \mathscr{B}_1\big) \big\vert \Psi },
\end{eqnarray}
where
\begin{equation}
    \mathscr{A}_x \equiv A^0_x - A^1_x,\qquad \mathscr{B}_y \equiv B^0_y - B^1_y.
\end{equation}
The quantity,
\begin{equation}
    \mathscr{C} \equiv \mathscr{A}_0 \big(\mathscr{B}_0 + \mathscr{B}_1\big) + \mathscr{A}_1 \big(\mathscr{B}_0 - \mathscr{B}_1\big),
\end{equation}
is the usual CHSH operator, satisfying the Tsirelson bound,
\begin{equation}\label{eq:TsirelsonC}
    \big\vert\!\braket{\mathscr{C}}\big\vert \leq 2\sqrt{2},
\end{equation}
and thus $\braket{\textrm{score}}\leq 1/\sqrt{2}$.

Recalling that the CHSH game awards +1 point for winning and -1 point for losing, the expected score per round is related to the probability of winning the round via
\begin{equation}\label{eq:CHSH-score-winloss}
    \braket{\textrm{score}} = P_\textrm{win} - P_\textrm{loss} = 2P_\textrm{win} - 1,
\end{equation}
where $P_\textrm{win}$ is the probability of winning the round and $P_\textrm{loss}=1-P_\textrm{win}$ is the probability of losing the round. Substituting Tsirelson's bound for the score and solving (\ref{eq:CHSH-score-winloss}) for the win probability yields
\begin{equation}
    P_\textrm{win} \leq \frac{1+\sqrt{2}}{2\sqrt{2}}.
\end{equation}

\section{Future Directions and Conclusion}\label{sec:Conclusion}

Our work demonstrates that treating the CHSH game as a causally local unistochastic process reproduces the Tsirelson bound for the CHSH game. In particular, our intuitively natural condition of causal locality threads a fine needle between violating the CHSH inequality up to the Tsirelson bound while preserving the no-signaling principle. 

Does causal locality imply that quantum systems whose dynamics satisfy the no-signaling condition are themselves local, despite some famous arguments to the contrary? The trouble with drawing this conclusion is that textbook quantum theory does not supply a way to define local causal influence at a microphysical level, as has been argued in \cite{SQCausalLocality}. The standard Copenhagen interpretation refrains from assigning any picture whatsoever to the microphysical goings-on of our systems. As such, it precludes any possibility of a causally local explanation underlying the behavior of the system from one measurement to another.

Causal locality relies on a different conceptual framework, one that, in contrast to standard quantum theory and the Copenhagen interpretation, does paint a picture of underlying laws governing the behavior of systems as they evolve from one configuration to another.  All of the game's actions can be formulated in terms of generically indivisible stochastic processes that occur entirely within a light cone that encompasses the entangling of qubit systems $Q$ and $R$, their transmission to Alice and Bob, Alice and Bob's subsequent local manipulations and measurements of $Q$ and $R$, and, ultimately, the tallying of the score by Charlie.

Our work implies that a violation of the Tsirelson bound requires either a violation of the principle of causal locality, and thus manifestly non-local dynamical laws, or requires laws that cannot be formulated in terms of unistochastic processes. Are there frameworks for dynamical laws that go beyond the indivisible stochastic systems formulated in \cite{SQCorrespondence}? We leave that question to future exploration.

Other approaches specialize from the very beginning to quantum systems that are explicitly formulated in terms of unitary time-evolution operators on Hilbert spaces \cite{Buhrman2005}. According to these alternative approaches, one needs to impose a technical condition to obtain violations of Tsirelson’s bound. This technical condition is that for a composite system consisting of localized subsystems at spacelike separation, the overall time-evolution operator should fail to tensor-factorize into unitary factors that individually encode the time evolution of each localized subsystem. One can show that such arrangements would lead to faster-than-light signaling. Our work, by contrast, does not assume that we are limiting ourselves to the standard framework of textbook quantum theory based on a Hilbert-space formulation, thus allowing us to draw broader implications in the context of causally local unistochastic systems. 

In developing the tools for our re-examination of the CHSH game, we explored the connection between division events and decoherence via the stochastic-quantum correspondence for systems that possess memory. Such systems may exhibit division events that produce a Markov-like washing away of information at times before the most recent division time, but marginalizing over their memory subsystem leaves the remaining subsystem with a non-Markov-like chain of conditional probabilities. More concretely, consider a system $A=SM$ with a subsystem $S$ and a memory $M$ that keeps track of the configurations $s_0,s_1,\ldots$ at the division events of the overall system. The effective dynamics of $S$ are non-Markovian in the sense that
\begin{equation}\begin{split}
    & p(s_t,s_{n},\ldots,s_1,s_0) \\
    &= p(s_t|s_n,\ldots,s_1,s_0) p(s_n|s_{n-1},\ldots,s_1,s_0)\cdots p(s_1|s_0) p(s_0),
\end{split}
\end{equation}
where $t>t_n >\cdots >t_1 > t_0$ and the times $t_0,t_1,\ldots,t_n \in \mathcal{T}^A_\textrm{div}$, where $\mathcal{T}^A_\textrm{div}$ is the set of division times for the system $A$. Note that we take time to be continuous, so the system evolves indivisibly between division times. Note, too, that $t$ is not assumed to be a member of the set of division times.

We conjecture that the quantum description of a non-Markovian conditional probability distribution is given by a case of the quantum comb or equivalent process-tensor constructions of \cite{QuantumComb:2008,QuantumCombClassical:2020,MilzModiQuantumStochastic2021},
\begin{equation}
    \label{eq:GeneralConditioningModel}
    p(s_t|s_n,s_{n-1},\ldots,s_0) = \textrm{Tr}\Big[\mathcal{P}_{s_t} \otimes \mathcal{P}_{s_n}\otimes\cdots \otimes \mathcal{P}_{s_0}\,\mathcal{E}^{(n)}_{t\from t_n}\big\{\mathcal{K}_n\{\mathcal{P}_{s_n}\otimes\cdots \otimes \mathcal{P}_{s_0}\}\big\}\Big],
\end{equation}
where
\begin{equation}
    \mathcal{K}_n\big\{\rho^{(n)}\big\} = \sum_{\mathclap{s'_0,\ldots,s'_n}} \mathcal{P}_{s'_n} \otimes \cdots\otimes \mathcal{P}_{s'_0}\,\rho^{(n)}\,\mathcal{P}_{s'_n} \otimes \cdots\otimes \mathcal{P}_{s'_0}
\end{equation}
acts as an $(n+1)$-time dephasing channel on an $(n+1)$-time density matrix $\rho^{(n)}$, and the time evolution of $S$ from $t_n\to t$ is represented by the evolution channel
\begin{equation}
    \mathcal{E}^{(n)}_{t\from t_n}\{\mathcal{P}_{s_n}\otimes\cdots \otimes \mathcal{P}_{s_0}\}= \mathcal{R}_{s_n,\ldots,s_0}(t)\otimes \mathcal{P}_{s_n}\otimes\cdots \otimes \mathcal{P}_{s_0},
\end{equation}
with
\begin{equation}
    \mathcal{P}_s \mathcal{R}_{s_n,\ldots,s_0}(t) \mathcal{P}_s = p(s|s_n,\ldots,s_0) \mathcal{P}_s.
\end{equation}
Note that the division times of the whole system $A$ do not reduce to division times for the subsystem $S$ because $S$ alone lacks the Markov-like property at these times. Nevertheless, joint and conditional probability distributions for $S$ can depend on the division times of $A$ in a non-Markovian manner. We thus refer to the division times of the whole system as effective conditioning times when describing subsystem $S$ alone.

Finally, it would be interesting to explore the structure of interference terms such as $\textrm{qCor}(t\from t'\from t_0)$ in (\ref{eq:Background:unistochasticFactors}) that arise from the discrepancy between unistochastic processes and their Markovian approximations. We believe that the nontrivial structure of such matrices may provide a route toward a novel proof of Tsirelson's bound and similar results, bypassing the need for explicit use of the quantum side of the stochastic-quantum correspondence.

\section*{Acknowledgements}

M.H. was supported by the UMass Dartmouth Marine and Undersea Technology (MUST)
Research Program sponsored by the Office of Naval Research (ONR) under grant N00014-22-1-
2012. D.K. was supported by the UMass Dartmouth MUST Research Program sponsored by the ONR under grant N00014-25-1-2315. We are also grateful for support from the American Institute of Mathematics.

\printbibliography

\end{document}